\documentclass[10pt]{article}
\usepackage[top=2cm,left=2cm,right=2cm]{geometry}
\usepackage[numbers,square]{natbib}

\usepackage[english]{babel}
\usepackage[latin1]{inputenc}
\usepackage{hyperref}
\usepackage{amssymb}
\usepackage{amsfonts}
\usepackage{amsmath}
\usepackage{fancyhdr}
\usepackage{epsfig}
\usepackage{graphics}
\usepackage{subfigure}
\usepackage{rotating}
\usepackage{lineno}
\usepackage{pstricks}
\usepackage{color}

            \let\d = \delta        
                  
                        \let\s = \sigma    \let\t = \tau

\newcommand{\muu}{\mu_{1}}
\newcommand{\nuu}{\nu_{1}}
\newcommand{\ku}{k_{1}}
\newcommand{\xu}{x_{1}}

\newcommand{\mud}{\mu_{2}}
\newcommand{\nud}{\nu_{2}}
\newcommand{\kd}{k_{2}}
\newcommand{\xd}{x_{2}}

\newcommand{\mut}{\mu_{3}}
\newcommand{\nut}{\nu_{3}}
\newcommand{\kt}{k_{3}}
\newcommand{\xt}{x_{3}}

\newcommand{\kq}{k_{4}}

\newcommand{\kc}{k_{5}}

\newcommand{\beq}{\begin{equation}}
\newcommand{\eeq}{\end{equation}}
\newcommand{\beqa}{\begin{eqnarray}}
\newcommand{\eeqa}{\end{eqnarray}}
\newcommand{\bea}{\begin{eqnarray}}
\newcommand{\eea}{\end{eqnarray}}

\newcommand{\nn}{\nonumber}
\newcommand{\spu}{\,\,\,}

\newcommand{\pd}{\partial}

\begin{document}
\begin{center}
\vspace{4.cm}
{\bf \large Conformal Trace Relations from the Dilaton Wess-Zumino Action \\}
\vspace{1cm}
{\bf Claudio Corian\`{o}, Luigi Delle Rose, Carlo Marzo and Mirko Serino}

\vspace{1cm}

{Dipartimento di Matematica e Fisica \\ 
Universit\`a del Salento \\ and \\ INFN Lecce, Via Arnesano 73100 Lecce, Italy\\}
\vspace{0.5cm}

\begin{abstract}

We use the method of Weyl-gauging in the determination of the Wess-Zumino conformal anomaly action to show that 
in any even ($d = 2 k$) dimensions all the hierarchy of correlation functions involving traces of the
energy-momentum tensor is determined in terms of those of lower orders, up to $2 k$. 
We work out explicitly the case $d=4$, and show that in this case in any conformal field theory 
only the first 4 traced correlators  are independent. 
All the remaining correlators are recursively generated by the first 4. 
The result is a consequence of the cocycle condition which defines the Wess-Zumino action  
and of the finite order of its dilaton interactions.

\end{abstract}

\end{center}

\newpage

\section{Introduction} 

The computation of correlation functions involving multiple traces of the stress-energy tensor in a generic conformal field theory 
is crucial for several applications, from the analysis of graviton vertices to studies of the ADS/CFT correspondence 
\cite{Henningson:1998gx}, being directly related to the conformal anomaly.  

We recall that in a generic field theory, defining the generating functional of the theory $\mathcal{W}$  as
\beq
\mathcal{W}[g] = \int \mathcal D \Phi\, e^{- \mathcal S}\, ,
\eeq
where $\mathcal{S}$ is the generic euclidean action depending on the set of all the quantum fields ($\Phi$) and on the 
background metric ($g$), the energy-momentum tensor (EMT) is given by
\beq \label{EMTvev}
\left\langle T^{\mu\nu}(x) \right\rangle 
= \frac{2}{\sqrt{g_x}}\, \frac{\delta \mathcal{W}[g]}{\delta g_{\mu\nu}(x)}
= \frac{2}{\sqrt{g_x}}\, \frac{\delta}{\delta g_{\mu\nu}(x)} \int\, \mathcal{D}\Phi\,  e^{-S}\, 
\eeq
and contains the response to the metric fluctuations. 
Here $g_x\equiv \left|g_{\mu\nu}(x)\right|$ is the determinant of the metric tensor.

For conformal field theories coupled to a background metric $g_{\mu\nu}(x)$ the anomaly condition takes the form
\beqa \label{TraceAnomaly}
g_{\mu\nu} \langle T^{\mu\nu} \rangle_g = \mathcal{A}[g]\, ,
\eeqa
where the presence of the subscript $g$ indicates a gravitational background. An anomalous relation of the form (\ref{TraceAnomaly}) holds in any even dimensions.
The anomaly functional for $d=4$ dimensions in the Dimensional Regularization (DR) scheme is defined by 
\beq
\mathcal{A}[g] = 
\beta_a\, \left( F - \frac{2}{3}\, \Box R \right)  + \beta_b\, G \, , \quad 
\beta_{a/b} = \sum_{I=S,F,V} n_I\,\beta_{a/b}(I)\, ,
\eeq

where $F$ is the squared four-dimensional Weyl tensor and $G$ the Euler density, both defined in the appendix.
The index $I$ runs over three kinds of fields, where $S$, $F$ and $V$ refer to scalars, fermions and abelian gauge bosons, 
determining their contribution to the anomaly through the anomaly coefficients $\beta_{a/b}$.
In general, multiple stress-energy tensor correlators can be defined in various ways, differing by contact terms \cite{Marzo:2012}.
These depend on the positions
of the $g^{-\frac{1}{2}}$ factors entering in the definition of the EMT respect to the functional derivatives.
We choose to define the Green function of $n$ EMT's in flat space in the completely symmetric fashion as 
\beq \label{NPF}
\langle T^{\mu_1\nu_1}(x_1)\ldots T^{\mu_n\nu_n}(x_n)\rangle 
\equiv
\frac{2^n}{\sqrt{g_{\xu}}\ldots \sqrt{g_{x_n}}}
\frac{\delta^n \mathcal{W}[g]}{\delta g_{\mu_1\nu_1}(\xu)\ldots \ldots\delta g_{\mu_n\nu_n}(x_n)}
\bigg|_{g_{\mu\nu}=\delta_{\mu\nu}} \, .
\eeq
It is also useful to introduce some notation to denote the functional derivatives 
with respect to the metric of generic functionals in the limit of a flat background
\beqa \label{funcder}
\left[f(x)\right]^{\muu\nuu\dots\mu_{n}\nu_{n}}(\xu,\dots,x_n) 
\equiv
\frac{\delta^n\, f(x)}{\delta g_{\mu_n\nu_n}(x_{n}) \, \ldots\, \delta g_{\muu\nuu}(\xu)}
\bigg|_{g_{\mu\nu}=\delta_{\mu\nu}} 
\eeqa
and the corresponding expression with traced indices
\beq
\left[f(x)\right]^{\muu\dots\mu_n}_{\spu\muu\dots\nu_n}\left(\xu,\xd,\dots,x_n\right)
\equiv \delta_{\muu\nuu}\dots\delta_{\mu_{n}\nu_{n}}\,
\left[f(x)\right]^{\muu\nuu\dots\mu_{n}\nu_{n}}\left(\xu,\dots,x_n\right)\, ,
\eeq
where the curved euclidean metric $g_{\mu\nu}$ is replaced by $\delta_{\mu\nu}$.

It is clear that, in any conformal field theory in even dimensions, the only object which plays a role in the
determination of the traces of these correlators is the anomaly functional, as one can realize by a direct computation. 
Specifically, from (\ref{TraceAnomaly}) one can derive trace identities for the $n$-point correlation functions.
In fact, in momentum space the entire hierarchy, which is generated by functional differentiation of (\ref{TraceAnomaly}),
takes the form
\beqa
\label{hier}
\left\langle T(\ku) \, \dots \, T(k_{n+1})\right\rangle
&=&
2^n\, \left[\sqrt{g}\, \mathcal A \right]^{\muu\dots\mu_n}_{\spu\muu\dots\nu_n}\left(\ku,\dots,k_{n+1}\right)
\nn \\
&&
-\, 2 \sum_{i=1}^{n} \left\langle T(\ku)\dots T(k_{i-1})T(k_{i+1})\dots T(k_{n+1}+k_i) \right\rangle \, .
\eeqa
In the expression above we have introduced the notation $T \equiv {T^\mu}_\mu$ to denote the trace of the EMT. All 
the momenta characterizing the vertex are taken as incoming, just as specified in the appendix.

The identity (\ref{hier}) relates a $n$-point correlator to correlators of order $n-1$, together with the completely 
traced derivatives of the anomaly functionals $\sqrt{g}\,F, \sqrt{g}\,G$ and $\sqrt{g}\,\square R$. 
For $\sqrt{g}\,F$, which is a conformal invariant, they are identically zero.
For $\sqrt{g}\,G$ these are nonvanishing at any arbitrary order $n \geq 3$, 
whereas $\sqrt{g}\,\Box R$ contributes also to the trace of the two-point function.
Therefore Eq.~(\ref{hier}) defines an open hierarchy, which involves on the right-hand-side all the derivatives of the anomaly 
functional.

\subsection{Effective actions and anomaly-induced actions} 

Our goal is to show that the structure of the hierarchy is entirely fixed just by the first four correlators,
using as  starting relation the cocycle condition satisfied by the Wess-Zumino anomaly-induced action. We recall that a Wess-Zumino 
action is constructed by solving the constraints coming from the conformal anomaly and differs from the effective action computed, 
for example, by using ordinary perturbation theory to integrate out the matter fields. 

For instance, direct computations of 
several correlators \cite{Giannotti:2008cv} \cite{Armillis:2009pq} have shown that these are in agreement with the expression 
predicted by the non-local anomaly action proposed by Riegert \cite{Riegert:1984kt}. 
In this respect the Wess-Zumino and the Riegert's action show significantly different features. 

For instance, a Wess-Zumino form of the non-local anomaly action is regained from Reigert's expression only at the cost of 
sacrificing covariance, by the choice of a fiducial metric \cite{AntMot:1991}. However, the Wess-Zumino action, which solves the 
anomaly constraint by using an extra field, the dilaton, plays a key role in order to extract information on some significant 
implications of the anomaly, as we are going to show.  

We will derive this action from the Weyl-gauging of the anomaly counterterms in DR \cite{Duff:1977}. 
We will be using the term {\em renormalized action} to denote the 
anomaly-induced action which is given by the sum of the Weyl-invariant (non anomalous) terms, denoted by 
$\Gamma_0$, and of the DR counterterms $\Gamma_{\textrm{Ct}}$ which one extracts in ordinary perturbation theory 
\cite{Duff:1977}. These are expressed in terms of the square of the Weyl tensor $F$  and of the Euler density $G$.
Explicitly 

\beq
\label{oneloop}
\Gamma_{\textrm{ren}}[g, \tau]=\Gamma_0[g,\tau] + \Gamma_{\textrm{Ct}}[g],
\eeq
where the dependence on the dilaton $\tau$ in $\Gamma_0$ is generated by the Weyl-gauging of diffeomorphism invariant functionals of 
the metric. This action correctly reproduces the anomaly, which is generated by the Weyl variation of  
$\Gamma_{\textrm{Ct}}$.  

Generically, the cocycle condition summarizes the response of the functional $\Gamma$ under a Weyl-gauging of the metric
\beq
\label{fet}
g_{\mu\nu} \to \hat{g}_{\mu\nu} = g_{\mu\nu}\, e^{- 2\, \Omega(x)},
\eeq
where $\Omega(x)\equiv \tau(x)/\Lambda$ defines the local Weyl scaling of the background metric, with $\tau(x)$ being the dilaton 
field and $\Lambda$ a conformal scale. The change in $g_{\mu\nu}$ due to a Weyl transformation is compensated by the shift of 
$\tau$, as we will specify below. 
In particular, defining the Weyl-gauged renormalized effective action
\beq
\hat\Gamma_{\textrm{ren}}[g,\tau] \equiv \Gamma_{0}[g,\tau] + \Gamma_{\textrm{Ct}}[\hat{g}]\, ,
\eeq
as we are going to show, the Wess-Zumino action will be identified from the relation
\beq
\label{coc}
\hat\Gamma_{\textrm{ren}}[g,\tau] = \Gamma_{\textrm{ren}}[g,\tau] - \Gamma_{\textrm{WZ}}[g,\tau].
\eeq
Here $\Gamma_{WZ}[g,\tau]$ is the Wess-Zumino action, whose Weyl variation equals the trace anomaly. 
 Notice that ${\hat{\Gamma}}_{\textrm{ren}}[{g,\tau}]$, as one can immediately 
realize, is Weyl invariant by construction, being a functional only of $\hat{g}$. We will come back to illustrate this 
point below. 
 
We will proceed with a rather general derivation of $\Gamma_{WZ}$ for $d=4$, which is quartic in the dilaton field $\tau$. 
In particular, we will determine the most general structure of the effective action in the DR scheme, with the inclusion of the finite counterterms in 
$\Gamma_{\textrm{Ct}}$.
The vanishing of all the dilaton interactions in $\Gamma_{WZ}$ above the 4-th order ones ensures that the hierarchy (\ref{hier}) can be 
expressed only in terms of 2-, 3- and 4-point functions of traced stress-energy tensors, proving our result. Our approach, as one can 
easily figure out, can be extended to any even spacetime dimensions, with the obvious change of the anomaly functional 
in (\ref{hier}), and can be used to infer the structure of the hierarchy of the traced correlators and of the 
Wess-Zumino action in an independent way.
 
This work is organized as follows. We start by reviewing the procedure of the Weyl-gauging for the determination of the 
Wess-Zumino anomaly action, which is identified directly from the gauging of the counterterms.
This point has been previously discussed in a cohomological context in \cite{MazMot:2001}. 
At this preliminary stage $\tau$ is just introduced as a compensator field, but the inclusion of kinetic terms for the same field, 
identified by the infinite sequence of Weyl invariants $\mathcal{J}_n(\hat{g})$ constructed from $\hat{g}_{\mu\nu}$, 
renders the dilaton dynamical.

These contributions introduce interactions of any order in $\tau$ but, 
as they are Weyl invariant, do not play any role in the identification of the constraints 
on the purely anomalous traces of the EMT's. As an explicit application of our method, we present the structure of the traced 5-point function of EMT's, computed by requiring the vanishing of the 5-dilaton interactions in $\Gamma_{WZ}$.  
For completeness, we have included in an appendix a brief review of the Noether method for the derivation of 
$\Gamma_{WZ}$, which is alternative to our approach and allows a useful cross-check of our results. A simpler application of our procedure in two dimensions can also be found in the appendix.

\section{Weyl-gauging and the Wess-Zumino action}

We briefly overview the method of Weyl-gauging (see \cite{Percacci:2012, Komargodski:2011, Freedman:2012}).

The metric tensor $g_{\mu\nu}(x)$, the vierbein $V_{a\,\rho}(x)$ 
and the fields $\Phi$ change under Weyl scalings according to
\beqa \label{WeylTransf}
g'_{\mu\nu}(x)     &=&  e^{2\, \sigma(x)}\, g_{\mu\nu}(x)\, ,  \nn \\
{V'}_{a\,\rho}(x)  &=&  e^{\sigma(x)}\, V_{a\,\rho}(x)\, , \nn \\
\Phi'(x)           &=&  e^{d_{\Phi}\, \sigma(x)}\, \Phi(x)\, ,
\eeqa
with $\sigma(x)$ being a dimensionless function parameterizing a local Weyl transformation.
Here $d_\Phi$ is the scaling dimension of the generic field $\Phi$, the latin suffix $a$ in $V_{a\rho}$ denotes the flat local index, 
while the greek indices are the curved indices of the spacetime manifold.
A way to build a Weyl invariant theory containing the fields in (\ref{WeylTransf}) is to apply the so called Weyl-gauging procedure, 
which consists in making the metric tensor, the vierbein and the fields $\Phi$ Weyl invariant through the substitution given in Eq. 
(\ref{fet}) together with
\bea \label{MetricGauge}
%
%
V_{a\,\rho}(x) &\rightarrow& \hat V_{a\,\rho}(x) \equiv  e^{- \frac{\tau(x)}{\Lambda}}\, V_{a\,\rho}(x) \, ,
\nn \\
\Phi(x)        &\rightarrow& \hat{\Phi}(x)       \equiv  e^{-d_{\Phi} \frac{\tau(x)}{\Lambda}}\, \Phi  \, 
\eea
and takes the form of field-enlarging transformations.

Under a Weyl scaling (\ref{WeylTransf}), the  dilaton $\tau$ shifts as a Goldstone mode
\beq \label{DilTransf}
\t'(x) = \tau(x) + \Lambda\, \s(x) \, .
\eeq  
In the case of the conformal anomaly as well as in a $U(1)$ gauge anomaly (cured with a Stuckelberg field, i.e. an axion), a field 
enlarging transformation, in the presence of a continuous symmetry, has no particular meaning unless the shift symmetry is broken. 
 
In the gauge anomaly case, Stuckelberg fields are introduced to restore the gauge invariance and acquire a physical meaning only in 
the presence of a Higgs-Stuckelberg mixing potential (Higgs-axion mixing) \cite{Coriano:2007fw}. They shift as Goldstone modes and 
couple to the anomaly, as in the case of $\tau$.

On the other hand, in the conformal case the anomaly does not need necessarily to be canceled and the presence both of a kinetic term 
for the dilaton and  of a linear coupling to the anomaly is sufficient to ensure the appearance of a massless but physical mode in 
the spectrum. Also in this case the dilaton is coupled by construction to the divergence of the broken dilatation current $J_D$
($\sim \tau \partial\cdot J_D=\tau\mathcal{A}$), as evident from the Noether construction reviewed in the appendix.

\subsection{The dynamical dilaton}

From (\ref{WeylTransf}) and (\ref{MetricGauge}) it also follows rather trivially that any diffeomorphism-invariant functional 
of the field-enlarged metric $\hat{g}_{\mu\nu}$ is Weyl invariant. 
According to the same equations, $\tau$ is a compensator field, 
which becomes dynamical as soon as we include some possible derivative terms for it. 
By applying the Weyl-gauging procedure to all the infinite set of diffeomorphism invariant functionals which can be built
out of the metric tensor and of increasing mass dimension, one can identify the homogenous terms of the anomaly action.
Beside, there will be the anomalous contributions, which are accounted for by the Wess-Zumino action. The latter can be added in 
order to identify a consistent anomaly action.
As we have already mentioned, the ambiguity intrinsic to the choice of the homogeneous terms is irrelevant in the determination of 
the structure of the hierarchy (\ref{hier}).  

The Weyl invariant terms may take the form of any scalar contraction of $\hat R_{\mu\nu\rho\sigma}$, $\hat R_{\mu\nu}$ and $\hat R$
and can be classified by their mass dimension. Typical examples are
\beq
\mathcal{J}_n\sim \frac{1}{\Lambda^{2(n-2)}}\int d^4 x \sqrt{\hat{g}}\hat{R}^n,
\eeq
and so forth, with the case $n=1$ describing the relevant operator in the infrared which reproduces the kinetic term of the dilaton. 
These terms can be included into $\Gamma_0[\hat{g}]\equiv \Gamma_0[g,\tau]$ which describes the non anomalous part of the renormalized action 
\beq
\Gamma_0[\hat{g}]\sim \sum_n \mathcal J_n[\hat{g}].
\eeq
The leading contribution to $\Gamma_0$ is the kinetic term for the dilaton, which can be obtained in two ways. The first method is to 
consider the Weyl-gauged Einstein-Hilbert term
\beqa
\int d^dx\, \sqrt{\hat g}\, \hat R 
&=& 
\int d^dx\, \sqrt{g}\, e^{\frac{(2-d)\,\tau}{\Lambda}}\, \bigg[
R - 2\, \left(d-1\right)\, \frac{\Box\tau}{\Lambda} +
\left(d-1\right)\,\left(d-2\right)\, \frac{\pd^\lambda\tau\,\pd_\lambda\tau}{\Lambda^2} \bigg] \nn \\
&=&
\int d^dx\, \sqrt{g}\, e^{\frac{(2-d)\,\tau}{\Lambda}}\, \bigg[R - 
\left(d-1\right)\,\left(d-2\right)\, \frac{\pd^\lambda\tau\,\pd_\lambda\tau}{\Lambda^2} \bigg]\, ,
\eeqa
with the inclusion of an appropriate normalization 
\beqa \label{EH}
\mathcal S^{(2)}_{\tau} = - \frac{\Lambda^{d-2}\, \left(d-2\right)}{8\,\left(d-1\right)}\, \int d^dx\, \sqrt{\hat g}\, \hat R  \, ,
\eeqa
which reverts the sign in front of the Einstein term. We recall that the extraction of a conformal factor ($\tilde{\sigma}$) 
from the Einstein-Hilbert term from a fiducial metric $\bar{g}_{\mu\nu}$ ($g_{\mu\nu}=\bar{g}_{\mu\nu}e^{\tilde{\sigma}}$) generates 
a kinetic term for ($\tilde{\sigma}$) which is ghost-like. In this case the non-local anomaly action, which in perturbation theory 
takes Riegert's form \cite{Riegert:1984kt}, can be rewritten in the Wess-Zumino form but at the cost of sacrificing covariance, due 
to the specific choice of the fiducial metric. 

We have attached to the action (\ref{EH})  a $"(2)"$ superscript to indicate that it reproduces terms which are 
quadratic in the derivatives. 

An alternative method consists in writing down the usual conformal invariant action for a scalar field $\chi$ 
in a curved background
\beqa \label{ScalarImprovedChi}
\mathcal S^{(2)}_{\chi} = \frac{1}{2}\, \int d^dx\, \sqrt{g}\, \bigg( 
g^{\mu\nu}\,\pd_\mu \chi\, \pd_\nu\chi - \frac{1}{4}\,\frac{d-2}{d-1}\, R\,\chi^2 \bigg) \, .
\eeqa
By the field redefinition $\chi\equiv\Lambda^{\frac{d-2}{2}}\, e^{-\frac{(d-2)\,\tau}{2\Lambda}}$ Eq.  (\ref{ScalarImprovedChi}) 
becomes
\beq \label{DilatonKinetic}
\mathcal S^{(2)}_{\tau} = \frac{\Lambda^{d-2}}{2}\, \int d^dx\, \sqrt{g}\, e^{-\frac{(d-2)\,\tau}{\Lambda}}\, \bigg( 
\frac{(d-2)^2}{4\,\Lambda^2}\, g^{\mu\nu}\,\pd_\mu \tau\, \pd_\nu\tau - \frac{1}{4}\, \frac{d-2}{d-1}\, R \bigg) \, ,
\eeq
which, for $d=4$, reduces to the familiar form
\beq \label{KinTau}
\mathcal S^{(2)}_{\tau} = \frac{1}{2}\, \int d^4 x\, \sqrt{g}\, e^{-\frac{2\,\tau}{\Lambda}}\, \bigg(
g^{\mu\nu}\,\pd_\mu \tau\, \pd_\nu\tau - \frac{\Lambda^2}{6}\, R \bigg)\, 
\eeq
and coincides with the previous expression (\ref{EH}), obtained from the formal Weyl invariant construction.

In four dimensions we can build the following possible subleading contributions (in $1/\Lambda$) to the effective action which, when gauged, 
can contribute to the fourth order dilaton action 
\beq
S^{(4)}_\tau = \int d^4x\, \sqrt{g}\, \bigg( \alpha\, R^{\mu\nu\rho\sigma}\, R_{\mu\nu\rho\sigma} + \beta\, R^{\mu\nu}\, R_{\mu\nu} 
                            + \gamma\, R^2 + \delta\, \Box R\bigg)\, .
\eeq
The fourth term ($\sim \Box R$) is just a total divergence, whereas two of the remaining three terms can be 
traded for the squared Weyl tensor $F$ and the Euler density $G$. 
As $\sqrt{g}\, F$ is Weyl invariant and $G$ is a topological term, neither of them contributes, 
when gauged according to (\ref{MetricGauge}), so that the only non vanishing four-derivative term
in the dilaton effective action in four dimensions is
\beq \label{UpToMarginal}
S^{(4)}_{\tau} = \gamma\, \int d^4x\, \sqrt{\hat{g}}\, \hat{R}^2 = 
\gamma\, \int d^4x\, \sqrt{g}\, 
\bigg[ R - 6\, \bigg( \frac{\Box\tau}{\Lambda} - \frac{\pd^\lambda\tau\,\pd_\lambda\tau}{\Lambda^2} \bigg) \bigg]^2 \, .
\eeq
with $\gamma$ a dimensionless constant.
If we also include a possible cosmological constant term, $\mathcal S^{(0)}_{\tau}$, we get the final form 
of the dilaton effective action in $d=4$ up to order four in the derivatives of the metric tensor
\beqa
\label{tot}
S_{\tau} &=& 
S^{(0)}_{\tau} + S^{(2)}_{\tau} + S^{(4)}_{\tau} + \dots =
\int d^4x\, \sqrt{\hat{g}}\, \bigg\{ \alpha
- \frac{\Lambda^{d-2}\, \left(d-2\right)}{8\,\left(d-1\right)}\, \hat{R} + \gamma\, \hat{R}^2\, 
\bigg\} + \dots\, ,
\eeqa
where the ellipsis refer to additional operators which are suppressed in $1/\Lambda$. 
In flat space ($g_{\mu\nu}\rightarrow \delta_{\mu\nu}$), (\ref{UpToMarginal}) becomes
\beq
\mathcal S_{\tau} = \int d^4x\, \bigg[ e^{-\frac{4\,\tau}{\Lambda}}\, \alpha + 
\frac{1}{2}\, e^{-\frac{2\,\tau}{\Lambda}} \, \pd^\lambda\tau\,\pd_\lambda\tau +
36\,\gamma\, \bigg( \frac{\Box\tau}{\Lambda} - \frac{\pd^\lambda\tau\,\pd_\lambda\tau}{\Lambda^2} \bigg) \bigg] + \dots
\eeq
where the ellipsis refer to higher dimensional contributions.
In general we can identify $\Gamma_0[\hat{g}]$ with $S_{\tau}$ as given in (\ref{tot}), thereby fixing the Weyl invariant 
contribution to $\Gamma_{\textrm{ren}}$.

\subsection{The counterterms} \label{Counterterms}

Having briefly reviewed the structure of the Weyl invariant operators in the dilaton effective action $\Gamma_{\textrm{ren}}$, we 
move to discuss the anomalous contributions. 

The standard approach followed in the derivation of the WZ anomaly action is the Noether method, in which $\tau$ is linearly coupled 
to the anomaly. Further terms are then introduced in order to correct for the non invariance under Weyl transformations of the 
anomaly functional itself. The approach, though very practical, does not make transparent the functional dependence of the 
Wess-Zumino action on the Weyl invariant metric $\hat{g}_{\mu\nu}$, which motivates our analysis. Here, instead, we proceed with a 
construction of the same effective action by applying the Weyl-gauging procedure to the renormalized effective action, which breaks 
scale invariance via the anomaly. The procedure allows to restore Weyl invariance by the field-enlarging transformation 
(\ref{MetricGauge}) and allows to extract the Wess-Zumino action up to a sign.

Following the discussion in \cite{Duff:1977}, we start by introducing the counterterm action
\beq\label{CounterAction4}
{\Gamma}_{\textrm{Ct}}[g] = 
- \frac{\mu^{-\epsilon}}{\epsilon}\int d^d x\, \sqrt{g}\, \bigg( \beta_a F + \beta_b G\bigg) \, ,  \quad \epsilon = 4 - d \, ,
\eeq
where $\mu$ is a regularization scale. It is this form of ${\Gamma}_{\textrm{Ct}}$, which is part of 
$\Gamma_{\textrm{ren}}$ (\ref{oneloop}) to induce the anomaly condition 
\beq
\label{qt}
\frac{2}{\sqrt{g}}g_{\mu\nu}\frac{\delta{\Gamma_{\textrm{ren}}}[g]}{\delta g_{\mu\nu}}\bigg|_{d\rightarrow 4} 
=\frac{2}{\sqrt{g}}g_{\mu\nu}\frac{\delta{\Gamma_{\textrm{Ct}}}[g]}{\delta g_{\mu\nu}}\bigg|_{d\rightarrow 4}= \mathcal{A}[g].
\eeq
In (\ref{qt}) we have exploited the Weyl invariance of the non anomalous action $\Gamma_0[g]$
\beq \label{BareActionInvariance}
g_{\mu\nu}\frac{\delta{\Gamma}_0[g]}{\delta g_{\mu\nu}}\bigg|_{d\rightarrow 4} = 0  \, , 
\eeq
with the anomaly generated entirely by the counterterm action ${\Gamma}_{\textrm{Ct}}[g]$. This follows from the well known relations
\beqa \label{NTracesCTF}
\frac{2}{\sqrt{g}}\, g_{\mu\nu}\, \frac{\delta}{\delta g_{\mu\nu}}\, \int d^d x\,\sqrt{g}\, F 
&=&
-\epsilon \, \left(F - \frac{2}{3}\, \Box R\right)\, , 
\\
\label{NTracesCTG}
\frac{2}{\sqrt{g}}\, g_{\mu\nu}\, \frac{\delta}{\delta g_{\mu\nu}}\, \int d^d x\, \sqrt{g}\, G 
&=& 
-\epsilon \, G \,,
\eeqa
which give 
\beq \label{TraceAnomaly4d}
\left\langle T \right\rangle = 
\frac{2}{\sqrt{g}}\, g_{\mu\nu}\, \frac{\delta{\Gamma}_{\textrm{Ct}}[g]}{\delta g_{\mu\nu}}\bigg|_{d\rightarrow 4} =  
\beta_a\, \left( F - \frac{2}{3}\, \Box R\right) + \beta_b\, G   \, .
\eeq
The $\Box R$ term in Eq.~(\ref{NTracesCTF}) is prescription dependent and can be avoided if the 
$F$-counterterm is chosen to be conformal invariant in $d$ dimensions, i.e. using the square $F_d$ of the Weyl tensor in $d$ 
dimensions (see Eq.~(\ref{SquaredWeyld})),
\beq \label{FdCounterterm}
{\Gamma}^{d}_{\textrm{Ct}}[g] = - \frac{\mu^{-\epsilon}}{\epsilon}\, \int d^d x\, \sqrt{g}\, \bigg( \beta_a F_d + \beta_b G\bigg)\, .
\eeq
In fact, expanding (\ref{FdCounterterm}) around $d=4$ and computing the $O(\epsilon)$ 
contribution to the vev of the traced EMT we find
\beqa
\int d^d x\,\sqrt{g}\, F_d 
&=& \int d^d x\,\sqrt{g}\, \bigg[ F - \epsilon\, \bigg( R^{\alpha\beta}R_{\alpha\beta} - \frac{5}{18}\, R^2 \bigg)
+ O\big(\epsilon^2\big) \bigg] \, , \\
\frac{2}{3}\, \Box R 
&=&
\frac{2}{\sqrt{g}}\, g_{\mu\nu}\, \frac{\delta}{\delta g_{\mu\nu}}\, 
\int d^4 x\, \sqrt{g}\, \left( R^{\alpha\beta} R_{\alpha\beta} - \frac{5}{18} R^2 \right) \, .
\eeqa
These formulae, combined with (\ref{NTracesCTF}), give
\beq \label{NTracesCT2}
\frac{2}{\sqrt{g}}\, g_{\mu\nu}\, \frac{\delta}{\delta g_{\mu\nu}}\, \int d^d x\,\sqrt{g}\, F_d
= - \epsilon\, F + O\big(\epsilon^2\big) \,
\eeq
in which the $\square R$ term is now absent.

In general, one may want to vary arbitrarily the coefficient in front of the $\Box R$ anomaly in (\ref{TraceAnomaly}).
This can be obtained by the inclusion of the counterterm
\beq \label{ll}
\beta_{\textrm{fin}}\, \int d^4x\, \sqrt{g}\, R^2\, ,
\eeq
where $\beta_{\textrm{fin}}$ is an arbitrary parameter, and the subscript $\textrm{fin}$ stands for "finite",
given that (\ref{ll}) is just a finite, prescription-dependent contribution.
In fact, the relation
\beq \label{LocalAnomaly}
\frac{2}{\sqrt{g}}\, g_{\mu\nu}\, \frac{\delta}{\delta g_{\mu\nu}}\, \int d^4 x\, \sqrt{g}\, R^2 = 12\, \Box R\, ,
\eeq
allows to modify at will the coefficient in front of $\square R$ in the anomaly functional. This is obtained by adding the finite 
contribution (\ref{ll}) to the action of the theory and by tuning appropriately the coefficient $\beta_{\textrm{fin}}$.
When (\ref{ll}) is present, the overall counterterm is
\beq
{\Gamma}_{\textrm{Ct}}[g] + \beta_{\textrm{fin}}\, \int d^4x\, \sqrt{g}\, R^2
\eeq
and the modified trace anomaly equation reads as
\beq
\left\langle T \right\rangle = 
\beta_a\, F + \beta_b\, G -\frac{2}{3}\, \bigg( \beta_a -18\, \beta_{\textrm{fin}} \bigg)\, \Box R \, .
\label{ModifiedTraceAnomalies}
\eeq
Nevertheless, the contribution (\ref{ll}) breaks the conformal symmetry of the theory. 
So the only modification of the effective action which modifies the coefficient of $\square R$ in the trace anomaly 
and is at the same time consistent with conformal symmetry is the replacement of $F$ with $F_d$, 
which removes the $\Box R$ anomaly altogether.

\subsection{Gauging the counterterms}\label{GaugeCount}

At this point we illustrate the practical implementation of the gauging procedure on the renormalized effective action.

It is natural to expand the gauged counterterms in a double power series with respect to $\epsilon = 4-d$
and $\kappa_{\Lambda}\equiv1/\Lambda$ around $(\epsilon , \kappa_\Lambda )=( 0,0)$. 
Their formal expansion is
\beq \label{PreGauging}
- \frac{1}{\epsilon}\,\int d^dx \, \sqrt{\hat{g}} \, \hat{F}\, (\hat{G}) = 
- \frac{1}{\epsilon}\, \int d^dx \sum_{i,j=0}^{\infty}\frac{1}{i!j!}\, \epsilon^i\,\left( \kappa_\Lambda\right)^j\,
\frac{\pd^{i+j}\left(\sqrt{\hat g}\, \hat F\, (\hat G)\, \right)}{\pd \epsilon^i\,\pd \kappa_\Lambda^{j}}\, .
\eeq
It is clear that only the $O(\epsilon)$ contributions are significant. 
On the other hand, the condition
\beq
\frac{\pd^{n} \left(\sqrt{\hat{g}} \, \hat{F}(\hat{G})\right)}{\pd\kappa_\Lambda^{n}} = O(\epsilon^2) \, , 
\quad n \geq 5 \, 
\eeq
is immediately found to hold, due to the absence of more than four dilatons in the gauged Riemann tensor
(see appendix \ref{Geometrical}). 
Beside, all the terms that are $O(1/\epsilon)$ in (\ref{PreGauging}) and are different from $F\, (G)$
are found to vanish after elementary integration by parts, so that we obtain the intermediate results
\beqa
\label{effe}
- \frac{\mu^{-\epsilon}}{\epsilon}\,\int d^dx \, \sqrt{\hat{g}} \, \hat{F} \, 
&=&
- \frac{\mu^{-\epsilon}}{\epsilon}\,\int d^dx \, \sqrt{g}\, F
+\int d^4x \, \sqrt{g}\, \bigg\{\frac{1}{\Lambda}\, 
\bigg( - \tau\, F - \frac{4}{3}\, R\,\Box\tau + 4\, R^{\alpha\beta}\, \nabla_\alpha\pd_\beta\tau \bigg)
\nn \\
&&
+\, \frac{2}{\Lambda^2}\, \bigg( 2\, R^{\alpha\beta}\,\pd_\alpha\tau\,\pd_\beta\tau - \frac{R}{3}\, \pd^\lambda\tau\,\pd_\lambda\tau
+ \left(\Box\tau\right)^2 - 2\, \nabla_\beta\pd_\alpha\tau\, \nabla^\beta\pd^\alpha\tau
\bigg) 
\nn \\
&&
-\, \frac{8}{\Lambda^3}\, \pd^\alpha\tau\,\pd^\beta\tau\, \nabla_\beta\pd_\alpha\tau
- \frac{2}{\Lambda^4}\, \left(\pd^\lambda\tau\,\pd_\lambda\tau\right)^2\, \bigg\}\, , 
\eea
\bea
\label{gi}
- \frac{\mu^{-\epsilon}}{\epsilon}\,\int d^dx \, \sqrt{\hat{g}} \, \hat{G} \,
&=&
- \frac{\mu^{-\epsilon}}{\epsilon}\,\int d^dx \, \sqrt{g}\, G
+ \int d^4x \, \sqrt{g}\, \bigg\{ \frac{1}{\Lambda}\, 
\bigg( - \tau\, G - 4\, R\, \Box \tau + 8\, R^{\alpha\beta}\, \nabla_\beta \pd_\alpha \tau \bigg)
\nn \\
&& \hspace{-5mm}
+\, \frac{2}{\Lambda^2}\, \bigg[ 2\, R\, \tau\,\Box\tau + R\, \pd^\lambda\tau\,\pd_\lambda\tau 
+ 4\, R^{\alpha\beta}\, \bigg(\pd_\alpha\tau\,\pd_\beta\tau - \tau\,\nabla_\beta \pd_\alpha\tau \bigg)
+ 6\, \left( \Box\tau \right)^2 - 6\, \nabla^\beta\pd^\alpha\tau\,\nabla_\beta\pd_\alpha\tau  \bigg]
\nn \\
&& \hspace{-5mm}
-\, \frac{4}{\Lambda^3}\, \bigg( 2\, R^{\alpha\beta}\,\tau\, \pd_\alpha\tau\, \pd_\beta\tau + 2\, \tau\, \left(\Box\tau\right)^2 +
5\, \pd^\lambda\tau\, \pd_\lambda\tau\, \Box\tau + 6\,\pd^\alpha\tau\,\pd^\beta\tau\,\nabla_\beta\pd_\alpha\tau - 
2\, \tau\, \nabla^\beta\pd^\alpha\tau\, \nabla_\beta\pd_\alpha\tau \bigg) 
\nn \\
&& \hspace{-5mm}
+\, \frac{2}{\Lambda^4}\, \bigg(
4\, \tau\, \pd^\lambda\tau\,\pd_\lambda\tau\, \Box\tau + 
3\, \left( \pd^\lambda\tau\,\pd_\lambda\tau \right)^2 + 
8\, \tau\, \pd^\alpha\tau\,\pd^\beta\tau\, \nabla_\beta\pd_\alpha \tau \bigg)\bigg\} \, .
\eeqa
Notice that the renormalization scale $\mu$ is not Weyl gauged due to the presence of the $d$-dimensional (rather than 4-dimensional) integration measure on the left-hand side of (\ref{effe}) and (\ref{gi}). 

The expressions above can be simplified using integrations by parts and the identity 
for the commutator of covariant derivatives of a vector
\beq
[\nabla_\mu,\nabla_\nu]\, v_\rho = {R^\lambda}_{\rho\mu\nu}\,v_\lambda  \, .
\eeq
After these manipulations we find that  the Weyl-gauging of the counterterms gives
\beqa
- \frac{\mu^{-\epsilon}}{\epsilon}\,\int d^dx \, \sqrt{\hat{g}} \, \hat{F} \, 
&=&  
- \frac{\mu^{-\epsilon}}{\epsilon}\, \int d^dx \, \sqrt{g} \, F +
\int d^4 x \, \sqrt{g}\,  \bigg[ - \frac{\tau}{\Lambda}\, \bigg( F - \frac{2}{3} \Box R \bigg)
- \frac{2}{\Lambda^2}\, \bigg( \frac{R}{3}\, \pd^\lambda\tau\, \pd_\lambda\tau + \left(\Box \tau \right)^2  \bigg)
\nn \\ 
&& \hspace{45mm}
+ \frac{4}{\Lambda^3}\, \pd^\lambda\tau\, \pd_\lambda\tau\,\Box\tau \, 
- \frac{2}{\Lambda^4}\, \left(\pd^\lambda\tau\, \pd_\lambda\tau\right)^2 \bigg] \, ,
\label{GaugingF} \\
- \frac{\mu^{-\epsilon}}{\epsilon}\,\int d^dx \, \sqrt{\hat{g}} \, \hat{G}
&=&
- \frac{\mu^{-\epsilon}}{\epsilon}\, \int d^dx \, \sqrt{g} \, G  + 
\int d^4 x \, \sqrt{g}\, \bigg[ -  \frac{\tau}{\Lambda}\, G 
+ \frac{4}{\Lambda^2}\, \left( R^{\alpha\beta} - \frac{R}{2}\,g^{\alpha\beta} \right)\, \pd_\alpha\tau\,\pd_\beta\tau
\nn \\
&& \hspace{45mm}
+ \, \frac{4}{\Lambda^3}\, \pd^\lambda\tau\, \pd_\lambda\tau\, \Box \tau
- \frac{2}{\Lambda^4}\, \left(\pd^\lambda\tau\, \pd_\lambda\tau \right)^2 \bigg]\, .
\label{GaugingG}
\eeqa  
Obviously, a Weyl variation applied to (\ref{GaugingF}) and (\ref{GaugingG}) gives zero by construction. It follows that the 
Wess-Zumino action can be extracted from (\ref{coc}) 

\beqa \label{ExtractWZ}
\Gamma_{WZ}[g,\tau] =\Gamma_{\textrm{ren}}[g,\tau] - \hat\Gamma_{\textrm{ren}}[g,\tau]
\eeqa
and thus takes the form 
\beqa
\Gamma_{WZ}[g,\tau] 
&=&
\int d^4x\, \sqrt{g}\, \bigg\{ \beta_a\, \bigg[ \frac{\tau}{\Lambda}\, \bigg( F - \frac{2}{3} \Box R \bigg) + 
\frac{2}{\Lambda^2}\, \bigg( \frac{R}{3}\, \pd^\lambda\tau\, \pd_\lambda\tau + \left( \Box \tau \right)^2 \bigg) - 
\frac{4}{\Lambda^3}\, \pd^\lambda\tau\,\pd_\lambda\tau\,\Box \tau + 
\frac{2}{\Lambda^4}\, \left( \pd^\lambda\tau\, \pd_\lambda\tau \right)^2 \bigg] \nn \\
&& 
\hspace{15mm}
+\, \beta_{b}\, \bigg[ \frac{\tau}{\Lambda}\,G  - 
\frac{4}{\Lambda^2}\, \bigg( R^{\alpha\beta} - \frac{R}{2}\,g^{\alpha\beta} \bigg)\, \pd_\alpha\tau\, \pd_\beta\tau -
\frac{4}{\Lambda^3} \, \pd^\lambda\tau\,\pd_\lambda\tau\,\Box \tau + 
\frac{2}{\Lambda^4}\, \left( \pd^\lambda\tau\, \pd_\lambda\tau \right)^2 \bigg]\bigg\} \, .
\label{Effective4d}
\eeqa
Notice that the ambiguity in the choice of the Weyl tensor discussed above - i.e. between F and $F_d$ - implies that no dilaton 
vertex is expected to emerge from the gauging of the $F_d$-counterterm, being the latter conformal invariant.
This is indeed the case and we find the relation
\beqa \label{GaugingFd}
&&  \hspace{-8 mm}
-\frac{\mu^{-\epsilon}}{\epsilon}\,\int d^d x \, \sqrt{\hat{g}} \, \hat{F}_d \, = 
-\frac{\mu^{-\epsilon}}{\epsilon}\,\int d^d x \, \sqrt{g} \, F_{d} 
- \int d^4 x \,\sqrt{g}\, \frac{\tau}{\Lambda}\, F\, ,
\eeqa
that modifies the structure of the Wess-Zumino action and it allows to eliminate all the terms proportional to $\beta_a$ 
in (\ref{Effective4d}) except for $\left(\tau/\Lambda\right)\, F$.

Finally, we remark that in the case in which a finite counterterm of the kind (\ref{ll}) is present,
the formulae of this section are modified according to the simple prescription (see Eq. (\ref{ModifiedTraceAnomalies})),
\beq \label{LocalToBeta}
\beta_a \rightarrow \beta_a - 18\, \beta_{\textrm{fin}}\, ,
\eeq
as it is possibile to render all the quantum effective action Weyl invariant. This is obtained, as discussed above, by the Weyl-gauging of the complete counterterm
\beq
\Gamma_{\textrm{Ct}}[g] = 
-\frac{\mu^{-\epsilon}}{\epsilon}\, \int d^dx\, \sqrt{g}\, \bigg( \beta_a\,F + \beta_b\,G \bigg) + 
\beta_{\textrm{fin}}\, \int d^4x\, \sqrt{g}\, R^2\, .
\eeq
In this case the compensating Wess-Zumino action for $\int d^4x\, \sqrt{g}\, R^2$ can be generated by the relation
\bea
\int d^4x\, \sqrt{\hat g}\, \hat{R}^2 
&=& 
\int d^4x\, \sqrt{g}\, R^2 +
18\, \int d^4 x \, \sqrt{g}\,  \bigg[ - \frac{2}{3}\, \frac{\tau}{\Lambda}\, \Box R
+ \frac{2}{\Lambda^2}\, \bigg( \frac{R}{3}\, \pd^\lambda\tau\, \pd_\lambda\tau + \left(\Box \tau \right)^2  \bigg)
\nn \\ 
&& \hspace{47mm}
- \frac{4}{\Lambda^3}\, \pd^\lambda\tau\, \pd_\lambda\tau\,\Box\tau \, 
+ \frac{2}{\Lambda^4}\, \left(\pd^\lambda\tau\, \pd_\lambda\tau\right)^2 \bigg] \, .
\eea
Comparing the result given above with (\ref{GaugingF}), Eq. (\ref{LocalToBeta}) follows immediately.

\section{Dilaton interactions and constraints from $\Gamma_{WZ}$}

Having extracted the structure of the Wess-Zumino action and thus of the anomaly-related dilaton interactions via
the Weyl-gauging of the effective action, we now follow a perturbative approach. 
We proceed with a Taylor expansion in $\kappa_{\Lambda}$ of the gauged metric which is given by 
\beq \label{SeriesInG}
\hat{g}_{\mu\nu} = g_{\mu\nu}\, e^{-2\,\kappa_{\Lambda}\tau} =
\bigg(\d_{\mu\nu} + \kappa\, h_{\mu\nu} \bigg)\, e^{-2\,\kappa_{\Lambda}\tau} =
\bigg(\delta_{\mu\nu} + \kappa\, h_{\mu\nu} \bigg)\,
\sum_{n=0}^{\infty} \frac{(-2)^n}{n!}\,(\kappa_{\Lambda}\,\tau)^n \, ,
\eeq
where $\kappa = \sqrt{16\,\pi\,G_{N}}$, with $G_{N}$ the Newton constant.
As we are considering only the dilaton contributions, we focus on the functional expansion of the
renormalized and Weyl-gauged effective action $\hat\Gamma_{\textrm{ren}}[g,\tau]$ with respect to $\kappa_{\Lambda}$.
This is easily done using the relation
\beq \label{CompositeDiff}
\frac{\pd\hat\Gamma_{\textrm{ren}}[g,\tau]}{\pd\kappa_{\Lambda}} = 
\int d^dy\, \frac{\delta\hat\Gamma_{\textrm{ren}}[g,\tau]}{\delta\hat{g}_{\mu\nu}(x)}
            \frac{\pd\hat{g}_{\mu\nu}(x)}{\pd\kappa_{\Lambda}}\, .
\eeq
Applying (\ref{CompositeDiff}) repeatedly and taking (\ref{SeriesInG}) into account, the perturbative series takes the form
\bea \label{Expansion}
\hat\Gamma_{\textrm{ren}}[g,\tau]
&=&
\Gamma_{\textrm{ren}}[g,\tau]
+\, \frac{1}{2!\,\Lambda^2}\, \int d^d \xu d^d \xd\, 
\frac{\delta^2\hat\Gamma_{\textrm{ren}}[g,\tau]}{\delta\hat{g}_{\muu\nuu}(\xu)\delta\hat{g}_{\mud\nud}(\xd)}
\frac{\pd \hat{g}_{\muu \nuu}(\xu)}{\pd \kappa_\Lambda}\frac{\pd \hat{g}_{\mud\nud}(\xd)}{\pd\kappa_\Lambda}
\nn \\
&& \hspace{10mm}
+\, \frac{1}{3!\, \Lambda^3}\, \bigg(\int d^d x_1 d^d \xd d^d \xt\,
\frac{\delta^3\hat\Gamma_{\textrm{ren}}[g,\tau]}
{\delta\hat{g}_{\muu\nuu}(\xu)\delta\hat{g}_{\mud\nud}(\xd)\delta\hat{g}_{\mut\nut}(\xt)}
\frac{\pd \hat{g}_{\muu \nuu}(\xu)}{\pd \kappa_\Lambda}\frac{\pd \hat{g}_{\mud \nud}(\xd)}{\pd \kappa_\Lambda}
\frac{\pd \hat{g}_{\mut \nut}(\xt)}{\pd \kappa_\Lambda}
\nn \\
&& \hspace{23mm}
+\, 3\, \int d^d \xu d^d \xd\, 
\frac{\delta^2\hat\Gamma_{\textrm{ren}}[g,\tau]}{\delta\hat{g}_{\muu\nuu}(\xu)\delta\hat{g}_{\mud\nud}(\xd)}
\frac{\pd^2 \hat{g}_{\muu\nuu}(\xu)}{\pd \kappa_\Lambda^2}\frac{\pd\hat{g}_{\mud\nud}(\xd)}{\pd \kappa_\Lambda} \bigg) +\ldots
\eea
As we are interested in the flat space limit of the dilaton action,
we write (\ref{Expansion}) by taking the limit of a conformally flat background metric 
$(\hat{g}_{\mu\nu}\rightarrow \hat \delta_{\mu\nu} \equiv\delta_{\mu\nu}\, e^{- 2\,\kappa_{\Lambda}\tau})$ obtaining
\bea \label{FinalExp}
\hat\Gamma_{\textrm{ren}}[\delta,\tau]
&=&
\Gamma_{\textrm{ren}}[\delta,\tau] + 
\frac{1}{2!\,\Lambda^2}\,  
\int d^d \xu d^d \xd\, \langle T(\xu) T(\xd)\rangle\, \tau(\xu)\tau(\xd)
\nn \\
&& 
-\, \frac{1}{3!\,\Lambda^3}\, \bigg[ \int d^d \xu d^d \xd d^d \xt\,\langle T(\xu) T(\xd) T(\xt)\rangle\, \tau(\xu)\tau(\xd)\tau(\xt)
\nn \\
&& \hspace{20mm}
+\, 6\,\int d^d \xu d^d \xd\, \langle T(\xu) T(\xd)\rangle\, (\tau(\xu))^2\tau(\xd) \bigg] + \ldots \, ,
\eea
where we have used Eq. (\ref{NPF}) in the definition of the EMT' s correlators and the obvious relation
\beq \label{MetricDil}
\frac{\pd^n \hat{g}_{\mu\nu}(x)}{\pd \kappa_{\Lambda}^n}\bigg|_{g_{\mu\nu}=\delta_{\mu\nu},\kappa_{\Lambda} = 0} = 
\left(-2\right)^n \, \left(\tau(x)\right)^n\, \delta_{\mu\nu}\, .
\eeq
From (\ref{FinalExp}) one may identify the expression of 
$\Gamma_{WZ}= \Gamma_{\textrm{ren}}[\delta,\tau] - \hat\Gamma_{\textrm{ren}}[\delta,\tau]$
written in terms of the traced n-point correlators of stress-energy tensors. 
This has to coincide with Eq. (\ref{Effective4d}) evaluated in the conformally flat limit and given by
\beq \label{FlatWZ}
- \Gamma_{WZ}[\delta,\tau] = 
- \int d^4x\,  \bigg[
\frac{2\,\beta_a}{\Lambda^2}\, \left( \Box \tau \right)^2
+ \left(\beta_a + \beta_b\right)\, \bigg( - \frac{4}{\Lambda^3}\, \pd^\lambda\tau\,\pd_\lambda\tau\,\Box \tau
+ \frac{2}{\Lambda^4}\, \left( \pd^\lambda\tau\, \pd_\lambda\tau \right)^2 \bigg) \bigg ] \, .
\eeq
At this point, a comparison between the dilaton vertices extracted from (\ref{FinalExp}) and (\ref{FlatWZ}) allows to establish a 
consistency condition between the first four of such vertices and a relation among the entire hierarchy of the traced correlators.

For this purpose we denote by $\mathcal I_n(\xu,\dots,x_n)$ the dilaton vertices obtained by functional differentiation of 
$\Gamma_{\textrm{ren}}[\hat{\delta}]$

\beq \label{FuncDiffWZ}
\mathcal I_{n}(\xu,\dots,x_n) = 
\frac{\delta^n \left(\Gamma_{\textrm{ren}}[\delta,\tau]-\hat\Gamma_{\textrm{ren}}[\delta,\tau]\right)}
{\delta\tau(\xu)\dots\delta\tau(x_n)} 
= -  \frac{\delta^n \Gamma_{WZ}[\delta,\tau]}{\delta\tau(\xu)\dots\delta\tau(x_n)}
\eeq
in coordinate space, which we can promptly transform to momentum space.
The expressions of such vertices for the first five orders in $\kappa_{\Lambda}$ are given by
\bea
{\mathcal I}_2(\ku,-\ku) 
&=& 
\frac{1}{\Lambda^2}\, \langle T(\ku) T(-\ku)\rangle \, ,
\nn \\
{\mathcal I}_3(\ku,\kd,\kt)
&=& 
-\frac{1}{\Lambda^3}\, \bigg[ 
\langle T(\ku) T(\kd) T(\kt) \rangle
+\, 2\, \bigg( \langle T(\ku) T(-\ku)\rangle + \langle T(\kd) T(-\kd)\rangle + \langle T(\kt) T(-\kt) \rangle \bigg) \bigg]\, ,
\nn \\ 
{\mathcal I}_4(\ku,\kd,\kt,\kq)
&=&
\frac{1}{\Lambda^4}\, \Biggl[
\langle T(\ku) T(\kd) T(\kt) T(\kq) \rangle
+\, 2\,\sum_{\mathcal T\left\{4,(k_{i_1},k_{i_2})\right\}}\langle T(k_{i_1}) T(k_{i_2}) T(-k_{i_1}-k_{i_2})\rangle
\nn \\
&& \hspace{6mm}
+\, 2\, \sum_{\mathcal T\left\{4,(k_{i_1},k_{i_2})\right\}} \langle T(k_{i_1}+k_{i_2}) T(-k_{i_1}-k_{i_2})\rangle
+ 4\, \sum_{i=1}^{4} \langle T(k_i) T(-k_i)\rangle \Biggr]\, ,
\nn \\
{\mathcal I}_5(\ku,\kd,\kt,\kq,\kc) 
&=&
- \frac{1}{\Lambda^5}\, \Biggl[
\langle T(\ku) T(\kd) T(\kt) T(\kq) T(\kc) \rangle 
\nn \\
&&
+\, 2\, \sum_{\mathcal T\left\{5,(k_{i_1},k_{i_2},k_{i_3})\right\}}
\langle T(k_{i_1}) T(k_{i_2}) T(k_{i_3}) T(-k_{i_1}-k_{i_2}-k_{i_3}) \rangle
\nn \\
&&
+ \,4\, \Biggl(
\sum_{\mathcal T\left\{5,(k_{i_1},k_{i_2})\right\}} \langle T(k_{i_1}) T(k_{i_2}) T(-k_{i_1}-k_{i_2}) \rangle 
\nn \\
&&
+\, \sum_{\mathcal T\left\{5,[(k_{i_1},k_{i_2}),(k_{i_3},k_{i_4})]\right\}}
\langle T(k_{i_1}+k_{i_2}) T(k_{i_3}+k_{i_4}) T(-k_{i_1}-k_{i_2}-k_{i_3}-k_{i_4}) \rangle \Biggr)
\nn \\
&&
+\, 8\, \Biggl(
\sum_{\mathcal T\left\{5,(k_{i_1},k_{i_2})\right\}} \langle T(k_{i_1}+k_{i_2}) T(-k_{i_1}-k_{i_2}) \rangle +
\sum_{i=1}^{5} \langle T(k_i) T(-k_i)\rangle \Biggr)
\Biggr]\, .
\label{DilIntStructure}
\eea
These can be extended to any higher order. We pause for a moment to clarify the notation used in 
(\ref{DilIntStructure}) for the organization of the momenta and the meaning of the symbol $\mathcal T$.

For example $\mathcal T\left\{4,(k_{i_1},k_{i_2})\right\} $ denotes the six pairs of distinct momenta in the case of the four point functions
\beq
\mathcal T\left\{4,(k_{i_1},k_{i_2})\right\} = 
\left\{(\ku,\kd),(\ku,\kt),(\ku,\kq),(\kd,\kt),(\kd,\kq),(\kt,\kq) \right\}   \, ,
\eeq
where we are combining the 4 momenta $k_1,...k_4$ into all the possible pairs, for a total of $\binom{4}{2}$ terms.
With five momenta $\left(\ku,\kd,\kt,\kq,\kc\right)$ the available pairs are
\bea
\mathcal T\left\{5,(k_{i_1},k_{i_2})\right\}
&=& 
\left\{(\ku,\kd),(\ku,\kt),(\ku,\kq),(\ku,\kc),(\kd,\kt),(\kd,\kq),(\kd,\kc),(\kt,\kq),(\kt,\kc),(\kq,\kc) \right\}  \, \nn\\
\eeqa
while the possible triples are 
\beqa
\mathcal T\left\{5,(k_{i_1},k_{i_2},k_{i_3})\right\}
&=& 
\left\{(\ku,\kd,\kt),(\ku,\kd,\kq),(\ku,\kd,\kc),(\ku,\kt,\kq),(\ku,\kt,\kc),
\right.
\nn \\
&&
\left.
\hspace{1mm}
(\ku,\kq,\kc),(\kd,\kt,\kq),(\kd,\kt,\kc),(\kd,\kq,\kc),(\kt,\kq,\kc) \right\}\, .
\eea
As we move to higher orders, the description of the momentum dependence gets slightly more involved and we need to distribute the external momenta 
into two pairs. The notation $\mathcal T\left\{5, [(k_{i_1},k_{i_2}),(k_{i_3},k_{i_4})]\right\}$ denotes the set of independent 
paired couples which can be generated out of $5$ momenta. 
Their number is $15$ and they are given by
\bea
\mathcal T\left\{5,[(k_{i_1},k_{i_2}),(k_{i_3},k_{i_4})]\right\}
&=& 
\left\{[(\ku,\kd),(\kt,\kq)],[(\ku,\kd),(\kt,\kc)],[(\ku,\kd),(\kq,\kc)] 
\right. 
\nn \\
&& \hspace{-55mm}
\left.     
[(\ku,\kt),(\kd,\kq)],[(\ku,\kt),(\kd,\kc)],[(\ku,\kt),(\kq,\kc)],[(\ku,\kq),(\kd,\kt)],
[(\ku,\kq),(\kd,\kc)],[(\ku,\kq),(\kt,\kc)], 
\right.
\nn \\
&& \hspace{-55mm}    
\left.
[(\ku,\kc),(\kd,\kt)],[(\ku,\kc),(\kd,\kq)],[(\ku,\kc),(\kt,\kq)],[(\kd,\kt),(\kq,\kc)],
[(\kd,\kq),(\kt,\kc)],[(\kd,\kc),(\kt,\kq)] \right\}\, .
\eea
It is obvious that a direct computation of $\mathcal{I}_2, \mathcal{I}_3$ and $\mathcal{I}_4$ 
from the anomaly action (\ref{FlatWZ}) allows to extract the explicit structure of these vertices in momentum space
\bea \label{DilatonInt}
{{\mathcal I}}_2(\ku,-\ku) 
&=&
-\frac{4}{\Lambda^2}\, \beta_a\, {\ku}^4\, , \nn \\
{{\mathcal I}}_3(\ku,\kd,\kt) 
&=& 
\frac{8}{\Lambda^3} \, \bigg(\beta_a + \beta_b \bigg)\, \bigg( 
\ku^2\, \kd\cdot\kt + \kd^2\, \ku\cdot\kt + \kt^2\, \ku\cdot\kd \bigg) 
\nn \\
{{\mathcal I}}_4(\ku,\kd,\kt,\kq)
&=&
- \frac{16}{\Lambda^4}\,\bigg(\beta_a + \beta_b \bigg)\, \bigg( 
\ku\cdot\kd\, \kt\cdot\kq + \ku\cdot\kt\, \kd\cdot\kq + \ku\cdot\kq\, \kd\cdot\kt  \bigg)\, ,
\eea
(with $k_{i}^n \equiv (k_i^2)^{n/2}$). These relations can be used together with (\ref{FinalExp}) in order to extract the 
structure of the 2- 3- and 4-point functions of the traced correlators, which are given by
\bea \label{BuildingBlocks}
\langle T(\ku) T(-\ku) \rangle 
&=& 
- 4\, \beta_a\, {\ku}^4 \, , 
\nn \\
\langle T(\ku) T(\kd) T(\kt) \rangle
&=& 
8 \bigg[ 
- \bigg( \beta_a+\beta_b \bigg)\,\bigg( f_{3}(\ku,\kd,\kt)+f_{3}(\kd,\ku,\kt)+f_{3}(\kt,\ku,\kd)\bigg) 
+        \beta_a\, \sum_{i=1}^{3} k_i^4 \bigg]\, ,
\nn \\
\langle T(\ku) T(\kd) T(\kt) T(\kq)\rangle 
&=&
8\, \bigg\{ 
6\, \bigg( \beta_a + \beta_b \bigg)\, \bigg[
\sum_{\mathcal T\left\{4,[(k_{i_1},k_{i_2}),(k_{i_3},k_{i_4})]\right\}} k_{i_i}\cdot k_{i_2}\, k_{i_3}\cdot k_{i_4}
\nn \\
&&
+\, f_{4}(\ku\,\kd,\kt,\kq) + f_{4}(\kd\,\ku,\kt,\kq) + f_{4}(\kt\,\ku,\kd,\kq) + f_{4}(\kq\,\ku,\kd,\kt) \bigg]
\nn \\
&&
-\, \beta_a\, 
\bigg( \sum_{\mathcal T\left\{4,(k_{i_1},k_{i_2})\right\}}(k_{i_1} + k_{i_2})^4 + 4\, \sum_{i=1}^{4} k_{i}^4 \bigg)
\bigg\}\, ,
\eea
where we have introduced the compact notation
\bea
f_{3}(k_a,k_b,k_c)
&=&
k_a^2\, k_b \cdot k_c \, ,
\nn \\
f_{4}(k_a,k_b,k_c,k_d)
&=&
k_a^2\, \left( k_b \cdot k_c + k_b \cdot k_d + k_c \cdot k_d \right)\, .
\eea
The third and fourth order results, in particular, were established in \cite{Marzo:2012}
via the explicit computation of the first three functional derivatives of the anomaly $\mathcal A[g]$
and exploiting recursively the hierarchical relations (\ref{hier}).

It is quite immediate to realize that the hierarchy in Eq. (\ref{hier}) can be entirely re-expressed in terms of the first four 
traced correlators. For this purpose, one has just to notice that $\Gamma_{WZ}[\hat{\delta}]$ is quartic in $\tau$, with  
$\mathcal{I}_n=0$, for $n\ge 5$. 
Therefore, for instance, the absence of vertices with 5 dilaton external lines, which sets $\mathcal{I}_5=0$, 
combined with the 4 fundamental traces in (\ref{BuildingBlocks}), are sufficient to completely fix the structure of the 5-point 
function, which takes the form 
\bea
&&
\langle T(\ku) T(\kd) T(\kt) T(\kq) T(\kc) \rangle =
16\, \Biggl\{
-24\,\bigg(\beta_a + \beta_b\bigg)\, \Biggl[
\sum_{\mathcal T\left\{5,[(k_{i_1},k_{i_2}),(k_{i_3},k_{i_4})]\right\}} k_{i_1}\cdot k_{i_2}\,k_{i_3}\cdot k_{i_4}
\nn \\
&& \hspace{-10mm}
+\, f_{5}(\ku,\kd,\kt,\kq,\kc) + f_{5}(\kd,\ku,\kt,\kq,\kc) + f_{5}(\kt,\ku,\kd,\kq,\kc) 
+   f_{5}(\kq,\ku,\kd,\kt,\kc) + f_{5}(\kc,\ku,\kd,\kt,\kq) \Biggr]
\nn \\
&& \hspace{-10mm}
+\, \beta_a\, \Biggl[
                     \sum_{\mathcal T\left\{5,(k_{i_1},k_{i_2},k_{i_3})\right\}} \left(k_{i_1} + k_{i_2} + k_{i_3} \right)^4
                     + 3\, \sum_{\mathcal T\left\{5,(k_{i_1},k_{i_2})\right\}} \left(k_{i_1} + k_{i_2} \right)^4
                     + 12\, \sum_{i=1}^{5} k_{i}^4 
\Biggr]
\Biggr\}\, ,
\label{5T}
\eea
where $f_5$ is defined as
\beq
f_{5}(k_a,k_b,k_c,k_d,k_e) = 
k_a^2\, \left( k_b \cdot k_c + k_b \cdot k_d + k_b \cdot k_e + k_c \cdot k_d + k_c \cdot k_e + k_d \cdot k_e  \right)\, .
\eeq
The construction that we have outlined can be extended to any arbitrary n-point function of traced stress-energy tensors. These 
relations can be checked in their consistency by a direct comparison with their equivalent expression obtained directly from the 
hierarchy (\ref{hier}). In general this requires the computation of functional derivatives of the anomaly functional $\mathcal{A}$ up 
to the relevant order. 
One can check by a direct computation using (\ref{hier}) the agreement with (\ref{5T}) up to the 5-th order. 
All the results given in this section can be easily generalized with the inclusion of a counterterm (\ref{ll}), 
using the prescription (\ref{LocalToBeta}), as discussed above. 

\section{Conclusions}

Our analysis has the simple goal of showing that the infinite hierarchy of fully traced correlation functions generated by the 
anomaly constraint in a generic conformal field theory in even dimensions has as fundamental building blocks only the first few 
correlators. For instance, in $d=4$ only correlators with $2, 3$ and $4$ traces are necessary to identify the entire hierarchy. 
This result can be simply derived from the structure of the Wess-Zumino action, which only contains dilaton 
interactions up to the quartic order. Non anomalous terms, which are homogenous under Weyl transformations and can be 
of arbitrarily higher orders in $\tau$, do not play any role in this construction. 
The Wess-Zumino action is determined, in general, by the Noether method, where the dilaton is coupled directly to the anomaly and
corrections are included in order to take care of the Weyl non-invariance of the functional. 
Alternatively, the same action is fixed by the cocycle condition, which shows that its functional dependence 
on the dilaton field takes place via the Weyl-gauging of the metric tensor. In our analysis we have introduced an expression of the 
anomaly-induced action in which the anomaly contribution is generated directly by the counterterms, evaluated in DR. This form of 
the action can be of significant help in the investigation of renormalization group flows in 
several dimensions, as we are going to show elsewhere.

\vspace{1cm}

\centerline{\bf \large Acknowledgements}

We thank Emil Mottola, Antonio Mariano and Pietro Colangelo for discussions.

\appendix %

\section{Conventions}\label{Geometrical}
 The definition of the Fourier transform for Eq. (\ref{NPF}) as well as for any n-point object is given by
\beq
\int \, d^d\xu\, \dots d^d x_n\, \left\langle T^{\muu\nuu}(\xu)\dots T^{\mu_n\nu_n}(x_n)\right\rangle \,
e^{-i(\ku\cdot \xu + \dots + k_n \cdot x_n)} = 
(2\pi)^d\,\delta^{(d)}\left( \sum_{i=1}^n k_i \right)\,\left\langle T^{\muu\nuu}(\ku)\dots T^{\mu_n\nu_n}(k_n)\right\rangle \, 
\label{NPFMom}
\eeq
with all the momenta incoming in the vertex. 
For the Riemann tensor we choose to adopt the sign convention
\beqa \label{Tensors}
{R^\lambda}_{\mu\kappa\nu}
&=&
\pd_\nu \Gamma^\lambda_{\mu\kappa} - \pd_\kappa \Gamma^\lambda_{\mu\nu}
+ \Gamma^\lambda_{\nu\eta}\Gamma^\eta_{\mu\kappa} - \Gamma^\lambda_{\kappa\eta}\Gamma^\eta_{\mu\nu}.
\eeqa
The traceless part of the Riemann tensor in $d$ dimensions is the Weyl tensor,
\beq \label{Weyl}
C_{\alpha\beta\gamma\delta} = R_{\alpha\beta\gamma\delta} -
\frac{1}{d-2}( g_{\alpha\gamma} \, R_{\beta\delta} - g_{\alpha\delta} \, R_{\beta\gamma}
             - g_{\beta\gamma}  \, R_{\alpha\delta} + g_{\beta\delta} \, R_{\alpha\gamma} ) +
\frac{R}{(d-1)(d-2)} \, ( g_{\alpha\gamma} \, g_{\beta\delta} - g_{\alpha\delta} \, g_{\beta\gamma})\, .
\eeq
Its square $F_d$ is given by 
\beq \label{SquaredWeyld}
F_d \equiv
C^{\alpha\beta\gamma\delta}C_{\alpha\beta\gamma\delta}
=
R^{\alpha\beta\gamma\delta}R_{\alpha\beta\gamma\delta} -\frac{4}{d-2}R^{\alpha\beta}R_{\alpha\beta}+\frac{2}{(d-1)(d-2)}R^2
\eeq
Its $d=4$ realization, called simply $F$, appears in the trace anomaly equation  (\ref{TraceAnomaly}).

The Euler density for $d=4$ is instead
\beq \label{Euler}
G = R^{\alpha\beta\gamma\delta}R_{\alpha\beta\gamma\delta} - 4\,R^{\alpha\beta}R_{\alpha\beta} + R^2\, .
\eeq
In the gauging of the counterterms we use the following relations 
\beqa
\hat {R^\mu}_{\nu\rho\sigma}
&=& 
{R^\mu}_{\nu\rho\sigma}
+ g_{\nu\rho}\,   \bigg( \frac{\nabla_{\sigma}\pd^\mu\tau}{\Lambda}  + \frac{\pd^\mu\tau\, \pd_\sigma\tau}{\Lambda^2} \bigg)
- g_{\nu\sigma}\, \bigg( \frac{\nabla_{\rho}\pd^\mu\tau}{\Lambda}    + \frac{\pd^\mu\tau\, \pd_\rho\tau}{\Lambda^2}   \bigg)
\nn \\
&& +\, 
{\delta^\mu}_\sigma\,\bigg( \frac{\nabla_{\rho}\pd_\nu\tau}{\Lambda}   + \frac{\pd_\nu\tau\,\pd_\rho\tau}{\Lambda^2} \bigg) -
{\delta^\mu}_\rho\,  \bigg( \frac{\nabla_{\sigma}\pd_\nu\tau}{\Lambda} + \frac{\pd_\nu\tau\,\pd_\sigma\tau}{\Lambda^2} \bigg) +
\bigg( {\delta^\mu}_\rho\, g_{\nu\sigma} - {\delta^\mu}_\sigma\, g_{\nu\rho} \bigg)\,
\frac{\pd^{\lambda}\tau\,\pd_	\lambda\tau}{\Lambda^2} \, , 
\nn \\
\hat R_{\mu\nu}
&=& 
R_{\mu\nu} - g_{\mu\nu}\, \bigg( \frac{\Box\tau}{\Lambda} - (d-2)\,\frac{\pd^\lambda\tau\,\pd_\lambda\tau}{\Lambda^2}\bigg) 
- (d-2)\, \bigg( \frac{\nabla_\mu \pd_\nu\tau}{\Lambda} + \frac{\pd_\mu\tau\,\pd_\nu\tau}{\Lambda^2} \bigg)\, ,
\nn \\
%
\hat R
&\equiv&
\hat g^{\mu\nu}\, \hat R_{\mu\nu} =
e^{\frac{2\,\tau}{\Lambda}}\bigg[ R - 2\, (d-1)\, \frac{\Box \tau}{\Lambda} 
+ (d-1)\,(d-2)\, \frac{\pd^\lambda\tau\,  \pd_\lambda \tau}{\Lambda^2} \bigg]\, .
\label{GaugeRiemann}
\eeqa

We use the variations of the Christoffel symbols
\beqa\label{deltaChristoffel}
\delta \Gamma^\alpha_{\beta\gamma}
&=&
\frac{1}{2}\,g^{\alpha\lambda}\big[ 
- \nabla_{\lambda}(\delta g_{\beta\gamma}) + \nabla_{\gamma}(\delta g_{\beta\lambda}) + \nabla_{\beta}(\delta g_{\gamma\lambda})
\big]\, ,
\nn\\
\nabla_\rho \delta\Gamma^\alpha_{\beta\gamma}
&=&
\frac{1}{2}\,g^{\alpha\lambda}\big[ - 
\nabla_{\rho}\nabla_{\lambda}(\delta g_{\beta\gamma}) + 
\nabla_{\rho}\nabla_{\gamma} (\delta g_{\beta\lambda}) + \nabla_{\rho}\nabla_{\beta}(\delta g_{\gamma\lambda}) \big]\, .
\eeqa
Specializing to the case of Weyl transformations, for which $ \delta_{W} g_{\mu\nu} = 2 \sigma g_{\mu\nu}$,
these formulae give the variations
\beqa \label{deltaWeylChristoffel}
\delta_{W} \Gamma^\alpha_{\beta\gamma}
&=&
- g_{\beta\gamma}\, \pd^\alpha \sigma + {\delta_\beta}^\alpha\, \pd_\gamma \sigma + {\delta_\gamma}^\alpha\, \pd_\beta\sigma
\quad \Rightarrow \quad \delta_{W} \Gamma^\alpha_{\alpha\gamma} = d\, \pd_\gamma \sigma \, , \nn \\
\nabla_\rho \delta_{W} \Gamma^\alpha_{\beta\gamma}
&=&
- g_{\beta\gamma}\, \nabla_\rho\pd^\alpha\sigma + {\delta_\beta}^\alpha\, \nabla_\rho\pd_\gamma\sigma
+ {\delta_\gamma}^\alpha\, \nabla_\rho\pd_\beta\sigma
\quad \Rightarrow \quad \delta_{W} \nabla_{\rho}\Gamma^\alpha_{\alpha\gamma} = d\, \nabla_\rho \pd_\gamma\sigma \, .
\eeqa
Using the Palatini identity
\beq \label{Palatini}
\delta {R^\alpha}_{\beta\gamma\rho} =
\nabla_{\rho}(\delta\Gamma^\alpha_{\beta\gamma}) - \nabla_{\gamma}(\delta\Gamma^\alpha_{\beta\rho})
\quad \Rightarrow \quad
\delta R_{\beta\rho} =
\nabla_{\rho} (\delta\Gamma^\lambda_{\beta\lambda}) - \nabla_{\lambda}(\delta\Gamma^\lambda_{\beta\rho})
\eeq
we obtain the expressions for the Weyl variations of the Riemann and Ricci tensors
\beqa \label{deltaWeylRiemann}
\delta_{W} {R^\alpha}_{\beta\gamma\rho} 
&=& 
  g_{\beta\rho}\, \nabla_{\gamma}\pd^\alpha\sigma 
- g_{\beta\gamma}\, \nabla_{\rho}\pd^\alpha \sigma
+ {\delta_{\gamma}}^{\alpha}\, \nabla_{\rho}\pd_\beta\sigma 
- {\delta_{\rho}}^{\alpha}\, \nabla_{\gamma}\pd_\beta\sigma \, , \nn \\
\delta_{W} R_{\beta\rho} 
&=& 
g_{\beta\rho}\, \Box \sigma + (d-2)\, \nabla_{\rho}\pd_\beta\sigma \, .
\eeqa
%

\section{ The Wess-Zumino action by the Noether method}\label{WessZumino}

The dilaton effective action that we have obtained by the Weyl-gauging of the counterterms can be recovered also through an iterative 
technique, that we briefly review.

In this second approach one begins by requiring that the variation of the dilaton effective action under the Weyl transformations
(\ref{WeylTransf}) and (\ref{DilTransf}) be equal to the anomaly
\bea
\delta \Gamma_{WZ}[g,\tau]
&=&
\int d^4x\, \sigma\, \bigg[ \beta_{a}\, \left( F - \frac{2}{3}\, \Box R \right) + \beta_b\, G \bigg]\, .
\label{VarWZ}
\eea
It is natural to start with the ansatz
\beq \label{ansatz1}
\Gamma^{(1)}_{WZ}[g,\tau] = 
\int d^4 x\, \sqrt{g}\, \frac{\tau}{\Lambda} \, \bigg[ \beta_{a}\, \left( F - \frac{2}{3}\, \Box R \right) + \beta_b\, G \bigg] \, .
\eeq
As $\sqrt{g}\, F$ is Weyl invariant, the variation of $\tau$ saturates the $F$-contribution in (\ref{VarWZ}).
But $\sqrt{g}\, G$ and $\sqrt{g}\, \Box R$ are not conformally invariant. Their variations under Weyl scalings
introduce additional terms that must be taken into account. The general strategy is to compute the infinitesimal variation
of these terms and to add terms which are quadratic in the derivatives of the dilaton and that cancel this extra contributions. 
But these, when Weyl-transformed, will generate additional terms which must be compensated in turn.
The iteration stops at the fourth order in the dilaton field.
We go through all the computation of the WZ action in some detail.
The piece of (\ref{ansatz1}) whose first Weyl variation is most easily worked out is the contribution $\sqrt{g}\,\tau\, \Box R$. 
We integrate it by parts twice and find that
\beqa
\delta_W \int d^4x\, \sqrt{g}\, \frac{\tau}{\Lambda}\, \Box R
&=& 
\delta_W \int d^4x\, \sqrt{g}\,g^{\mu\nu}\,g^{\rho\sigma}\, R_{\mu\nu}\, \frac{1}{\Lambda}\, \nabla_\sigma\pd_\rho\tau\, , 
\nn \\
&=&
\int d^4x\, \sqrt{g} \bigg( \Box\sigma\, R - \frac{1}{\Lambda}\, g^{\rho\sigma}\,\delta_W \Gamma^\lambda_{\rho\sigma}\, 
\pd_\lambda\tau\, R + \frac{1}{\Lambda}\,\Box \tau\, g^{\mu\nu}\, \delta_W R_{\mu\nu}  \bigg)\, .
\eeqa
Using (\ref{deltaWeylChristoffel}) and (\ref{deltaWeylRiemann}) this turns into
\beq \label{FirstWeylBoxR}
\delta_W \int d^4x\, \sqrt{g}\, \frac{\tau}{\Lambda}\, \Box R = 
\int d^4 x\, \sqrt{g}\, \bigg( \sigma\, \Box R + \frac{2}{\Lambda}\, R\, \pd_\lambda\tau\, \pd^\lambda\sigma 
+ \frac{6}{\Lambda}\, \Box\tau\Box\sigma \bigg) \, .
\eeq

Now we perform an infinitesimal Weyl variation of the contribution $\sqrt{g}\,\tau\,G$ and we find
\bea
\delta_{W} \int d^4x\, \sqrt{g}\, \frac{\tau}{\Lambda}\, G
&=& 
\int d^4 x\, \sqrt{g}\, \bigg\{ \sigma\, G  +
\frac{2}{\Lambda}\,\tau \bigg[ {R_{\alpha}}^{\beta\gamma\delta}\delta_W {R^{\alpha}}_{\beta\gamma\delta}
- \left(4\,R^{\alpha\beta} - g^{\alpha\beta}R \right) \delta_W R_{\alpha\beta} \bigg] \bigg\}\, .
\eea
After using the algebraic symmetries of the Riemann and Ricci tensors and relabeling the indices we obtain
\beqa
\delta_{W} \int d^4x\, \sqrt{g}\, \frac{\tau}{\Lambda} \, G
&=& 
\int d^4 x\, \sqrt{g}\, \bigg\{ 
\sigma\, G + \frac{\tau}{\Lambda}\, \bigg[ 4\,R^{\lambda\gamma\beta\alpha}
- 4\,\bigg( g^{\alpha\gamma}R^{\beta\lambda} + g^{\beta\lambda} R^{\alpha\gamma} - 2\,g^{\lambda\gamma}\,R^{\alpha\beta} \bigg) \nn \\
&& \hspace{35mm} 
+\, 2\, \bigg( g^{\alpha\gamma}g^{\beta\lambda} - g^{\alpha\beta}g^{\gamma\lambda} \bigg)\, R \bigg]\,
\nabla_\lambda\nabla_\beta(\delta_W g_{\alpha\gamma}) 
\bigg\}\, . \nn \\
\eeqa
Now we set $\delta_W g_{\alpha\gamma} = 2\, \sigma g_{\alpha\gamma}$ and after a double integration by parts we obtain
\beq \label{FirstWeylG}
\delta_{W} \int d^4x\, \sqrt{g}\, \frac{\tau}{\Lambda} \, G = 
\int d^4 x\, \sqrt{g}\, \bigg\{ \sigma\, G
+ \frac{8}{\Lambda}\,\bigg[ \bigg( R^{\alpha\beta}- \frac{1}{2}\, g^{\alpha\beta}\,R \bigg)\, \pd_\alpha\sigma\, \pd_\beta\tau
\bigg] \bigg\} \, .
\eeq
Combined together, Eqs. (\ref{FirstWeylBoxR}) and (\ref{FirstWeylG}) give the Weyl variation of the first ansatz (\ref{ansatz1})
\beqa \label{Var1}
\delta_{W} \Gamma^{(1)}_{WZ}[g,\tau] 
&=&
\int d^4 x\, \sqrt{g}\, \bigg\{
\sigma\, \bigg[\beta_a\, \left(F - \frac{2}{3}\, \Box R  \right) + \beta_b\, G \bigg]
\nn \\
&& \hspace{14mm}
+ \frac{1}{\Lambda}\, \bigg[ \beta_a\, 
\left( \frac{4}{3}\,R\, \pd^\lambda\tau\, \pd_\lambda\sigma + 4\, \Box \tau\, \Box \sigma \right)
+\, 8\, \beta_b\, \pd_\alpha\sigma\, \pd_\beta\tau\, \left( R^{\alpha\beta} - \frac{g^{\alpha\beta}}{2}\, R  \right) \bigg]
\bigg\}\, .
\eeqa
In order to cancel second line in the integrand in (\ref{Var1}) we correct with the second ansatz
\beqa \label{ansatz2}
\Gamma^{(2)}_{WZ}[g,\tau] = \Gamma^{(1)}_{WZ}[g,\tau]
+ \frac{1}{\Lambda^2}\,\int d^4 x\, \sqrt{g}\, \bigg\{
\beta_a\, \bigg(\frac{2}{3}\,R\, \pd^\lambda\tau\, \pd_\lambda\tau + 2\, \left(\Box\tau\right)^2 \bigg)
- 4\, \beta_b\, \bigg( R^{\alpha\beta} - \frac{g^{\alpha\beta}}{2}\, R \bigg)\, \pd_\alpha\tau\, \pd_\beta\tau \bigg\}\, . \nn \\
\eeqa
We then find that the variation of this second ansatz is given by
\beqa \label{Var2}
\delta_{W} \Gamma^{(2)}_{WZ}[g,\tau] 
&=&
\int d^4 x\, \sqrt{g}\, \bigg\{
\sigma\, \bigg[\beta_a\, \left(F - \frac{2}{3}\, \Box R  \right) + \beta_b\, G \bigg] 
+ \frac{1}{\Lambda^2}\, \bigg[ \beta_a\, \left( 8\, \Box \tau\, \pd^\lambda\tau\, \pd_\lambda\sigma
+ 4\, \pd^\lambda\tau\, \pd_\lambda\tau\, \Box \sigma \right) 
\nn \\
&& \hspace{16mm}
+\, 8\, \beta_b\, \left( \pd^\lambda\tau\,\pd_\lambda\tau \,\Box \sigma 
- \pd_\alpha\tau\,\pd_\beta\tau\, \nabla^\beta\pd^\alpha\sigma \right) 
\bigg]
\bigg\}\, .
\eeqa
It is then necessary to compensate for terms which are cubic in the dilaton. 
The structure of the spurious contributions in (\ref{Var2}) suggests the third ansatz
\beqa \label{ansatz3}
\Gamma^{(3)}_{WZ}[g,\tau] = \Gamma^{(2)}_{WZ}[g,\tau]
- \frac{4}{\Lambda^3}\,\int d^4 x\, \sqrt{g}\,
\left( \beta_a + \beta_b \right)\, \pd^\lambda\tau\,\pd_\lambda\tau\,\Box \tau \, .
\eeqa
The variation of (\ref{ansatz3}) is
\beqa \label{Var3}
\delta_{W} \Gamma^{(3)}_{WZ}[g,\tau] 
&=&
\int d^4 x\, \sqrt{g}\, \bigg\{ 
\sigma\, \bigg[\beta_a\, \left(F - \frac{2}{3}\, \Box R  \right) + \beta_b\, G \bigg]
+ \frac{4}{\Lambda^2}\, \beta_b\, \left( 2\, \nabla_\beta\pd_\alpha\tau\, \pd^\beta\tau\, \pd^\alpha\sigma
+ \pd^\lambda\tau\,\pd_\lambda\tau\, \Box \sigma \right) \nn \\
&& \hspace{22mm}
-\, \frac{8}{\Lambda^3}\, \left(\beta_a + \beta_b\right)\, \pd^\alpha\tau\,\pd_\alpha\tau\, \pd^\beta\tau\, \pd_\beta\sigma
\bigg\} \nn \\
&=&
\int d^4 x\, \sqrt{g}\, \bigg\{
\sigma\, \bigg[\beta_a\, \left(F - \frac{2}{3}\, \Box R  \right) + \beta_b\, G \bigg]
-\, \frac{8}{\Lambda^3}\, \left(\beta_a + \beta_b\right)\,
\pd^\alpha\tau\,\pd_\alpha\tau\, \pd^\beta\tau\,\pd_\beta\sigma \bigg\}\, ,
\eeqa
which finally allows to infer the structure of the complete WZ action 
\bea \label{ansatz4}
\Gamma_{WZ}[g,\tau] = \Gamma^{(3)}_{WZ}[g,\tau] 
+ \frac{2}{\Lambda^4}\, \int d^4 x\, \sqrt{g}\, \left( \beta_a + \beta_b \right)\,
\left(\pd^\lambda\tau\, \pd_\lambda\tau\right)^2 \, .
\eea
Eq. (\ref{ansatz4}) coincides with (\ref{Effective4d}) and it is easy to check that no more terms are needed to ensure
that (\ref{VarWZ}) holds.
%

\section{The recursive relation in 2 dimensions}

For general even dimensions $d = 2\, k$, the trace anomaly of a conformal field theory 
is always a functional of order $d$ in the derivatives of the metric tensor.
This implies, as clear from the Weyl-gauging procedure applied to the counterterms discussed in section \ref{GaugeCount},
that no dilaton interaction of order higher than $d$ can appear in the Wess-Zumino action.
Taking into account (\ref{FinalExp}), which holds in all dimensions, we conclude that, for any even $d$,
all the traced correlators with more than $d$ insertions of EMT's are related to lower order Green functions giving
\beq \label{NoInt}
\mathcal I_{n} = 0\, , \quad n \ge d\, .
\eeq
We present here the application of the analysis discussed in the main text to the simpler case of $d=2$.

The equation of the trace anomaly in two dimensions is
\beq \label{TraceAnomaly2D}
\langle T \rangle = - \frac{c}{24\,\pi}\, R\, ,
\eeq
where $c = n_s + n_f$, with $n_s$ and $n_f$ being the numbers of free scalar and fermion fields respectively.\\
The counterterm in DR is given by
\beq\label{Counterterm2D}
\Gamma_{\textrm{Ct}}[g] = - \frac{\mu^{\epsilon}}{\epsilon}\, \frac{c}{24\,\pi} \,\int d^d x\, \sqrt{g}\, R\, , 
\quad \epsilon = d - 2\, .
\eeq
The derivation of the trace anomaly from the counterterm is completely analogous to the case $d=4$, dicussed in \ref{Counterterms}.
From (\ref{qt}), (\ref{BareActionInvariance}) and elementary manipulations we obtain
\beq
g_{\mu\nu}\, \langle T^{\mu\nu} \rangle 
= \frac{2}{\sqrt{g}}\, g_{\mu\nu}\, \frac{\delta\Gamma_{\textrm{Ct}}[g]}{\delta g_{\mu\nu}}\bigg|_{d\rightarrow 2}
= - \frac{c}{24\,\pi}\, R   \, .
\eeq
We repeat the Weyl-gauging of section \ref{GaugeCount} for the integral of the scalar curvature 
in DR and find
\beq \label{GaugeCT2D}
-\frac{\mu^{\epsilon}}{\epsilon}\, \int d^dx \, \sqrt{\hat g}\, \hat R =
- \frac{\mu^{\epsilon}}{\epsilon}\, \int d^dx \, \sqrt{g} \, R + 
\int d^2 x \, \sqrt{g}\, \left[ \frac{\tau}{\Lambda}\, R + \frac{1}{\Lambda^2} \pd_\alpha\tau\, \pd^\alpha\tau \right]\, .
\eeq
The second term in (\ref{GaugeCT2D}) is, modulo a constant, the Wess-Zumino action in $2$ dimensions,
\beq
\Gamma_{WZ}[g,\tau] = - \frac{c}{24\,\pi}\, \int d^2x\, \sqrt{g}\, \left[
\frac{\tau}{\Lambda}\, R  + \frac{1}{\Lambda^2} \pd_\alpha\tau\, \pd^\alpha\tau 
\right] \, ,
\eeq
from which we can extract the 2-dilaton amplitude according to (\ref{FuncDiffWZ}). 
Requiring (\ref{NoInt}) to hold for $n \geq 3$, and using the relations in (\ref{DilIntStructure}) we find the expression
\beq
\mathcal I_{2}(\ku\,-\ku) = \frac{1}{\Lambda^2}\, \langle  T(\ku) T(-\ku) \rangle = \frac{c}{12\,\pi}\, \ku^2
\eeq
This 2-dilaton vertex is the only one which is non vanishing. Inverting the remaining relations we obtain
\bea
\langle T(\ku) T(\kd) T(\kt) \rangle 
&=& 
- \frac{c}{6\,\pi}\, \left( \ku^2 + \kd^2 + \kt^2 \right)\, ,
\nn \\
\langle T(\ku) T(\kd) T(\kt) T(\kq) \rangle 
&=&
\frac{c}{\pi}\, \left( \ku^2 + \kd^2 + \kt^2 + \kq^2 \right)\, , \nn \\
\langle T(\ku) T(\kd) T(\kt) T(\kq) \rangle 
&=&
-\frac{8\,c}{\pi}\, \left( \ku^2 + \kd^2 + \kt^2 + \kq^2 + \kc^2 \right)\, 
\eea
and so on for any Green function of higher order.
Also in this case we have cross-checked our results by comparing those obtained with this algorithm with the 
canonical ones obtained by functional differentiation of the anomaly (\ref{TraceAnomaly2D}).


\end{document}